\documentclass[]{aa}
\usepackage{psfig}
\usepackage{natbib}
\usepackage{epsf}
\usepackage{graphicx}

\bibpunct{(}{)}{;}{a}{}{,}

\begin{document}

\def\eps{\varepsilon}
\def\aap{A\&A}
\def\AaA{A\&A}
\def\apj{ApJ}
\def\ApJ{ApJ}
\def\ApJS{ApJS}
\def\apjl{ApJL}
\def\ARAA{ARAA}
\def\MNRAS{MNRAS}
\def\mnras{MNRAS}
\def\aj{AJ}
\def\AJ{AJ}
\def\nat{Nature}
\def\aaps{A\&A Supp.}
\def\me{m_\e}
\def\lesssim{\mathrel{\hbox{\rlap{\hbox{\lower4pt\hbox{$\sim$}}}\hbox{$<$}}}}
\def\gtrsim{\mathrel{\hbox{\rlap{\hbox{\lower4pt\hbox{$\sim$}}}\hbox{$>$}}}}
\def\cmb{{\rm cmb}}
\def\a{\alpha}
\def\e{{\rm e}}
\def\o{{0}}
\def\p{{\rm p}}
\def\me{m_{\rm e}}
\def\gh{{\rm gh}}
\def\cl{{\rm cl}}
\def\rg{{\rm rg}}
\def\th{{\rm th}}
\def\rel{{\rm rel}}
\def\jet{{\rm jet}}
\def\power{{\rm power}}
\def\aC{{a_{\rm C}}}
\def\ab{{a_{\rm b}}}
\def\as{{a_{\rm s}}}
\newcommand{\hq}{\hbar}
\newcommand{\epsB}{\varepsilon_{\rm B}}
\newcommand{\epsCMB}{\varepsilon_{\rm cmb}}

\def\CR{{\rm cr}}
\def\gal{{\rm gal}}
\def\gas{{\rm gas}}

\def\del#1{{}}

\def\C#1{#1}

\title{Trouble for cluster parameter estimation from blind SZ surveys? } 
\titlerunning{Trouble for cluster parameter estimation from blind SZ
  surveys?}
\author{Nabila Aghanim\inst{1}, Steen H. Hansen\inst{2},
\and Guilaine Lagache\inst{1}} 
\authorrunning{Aghanim, Hansen \& Lagache} 
\institute{IAS-CNRS, B\^atiment 121, Universit\'e Paris Sud 91405
Orsay, France \and University of Zurich, Winterthurerstrasse 190, 8057
Zurich, Switzerland} 
\date{}

\abstract{ The Sunyaev-Zel'dovich (SZ) effect of galaxy clusters is a
tool to measure three quantities: Compton parameter, electron
temperature, and cluster peculiar velocity. However, a major
problem is non-removed contamination by astrophysical sources that
emit in the SZ frequencies. This includes interstellar dust emission,
infra-red (IR) galaxies, and radio sources in addition to primary
Cosmic Microwave Background (CMB) anisotropies. The three former
contaminations induce systematic shifts in the three SZ
parameters. In this study, we carefully estimated, both for a
large beam experiment (namely Planck Surveyor) and a small beam
experiment (ACT-like), the systematic errors that result if a fraction
of the expected levels of emission from dust, IR galaxies, and radio
sources remains non-removed. We found that the interstellar dust
emission is not a major contaminant for the SZ measurement.
Unfortunately, the IR and radio source-induced systematic errors may
be extremely large. In particular the intra-cluster temperature
and peculiar velocity will be determined inaccurately for Planck
and ACT-like experiments, if only the frequency dependences are used
for the cleaning. The Compton parameter is also affected by the
astrophysical contaminations. The systematic errors in this case were a
factor of 2 to 5 times larger than the expected statistical error-bar
for Planck.  For the ACT-like experiment, the statistical error-bars
were larger than in the case of Planck by a factor of about 5, and
therefore the systematic shifts remain within about $50\%$ of the
statistical errors. We have thus shown that the systematic
errors due to contaminating astrophysical emissions can be
significantly larger than the statistical errors, which implies that
future SZ surveys aiming at measuring cluster temperatures and
peculiar velocities will not be able to do so on their own without
including additional information like cluster shapes or follow-up
observations.  }

\maketitle

\section{Introduction\label{sec:intro}}

The well-studied Sunyaev-Zel'dovich (SZ) effect \cite[]{sz72,sz80} 
has been 
observed towards a few tens of known galaxy clusters (see reviews by
\cite{carlstrom2002} and Birkinshaw 1999). It is a
powerful tool for cosmological and cluster studies, and when combined
with other observations (X-rays, optical, lensing) the SZ effect
allows us to measure cosmological parameters such as the Hubble
constant and matter density in the universe $\Omega_{\mathrm m}$
(e.g. \cite{myers97}, \cite{grego2001}, \cite{reese2002},
\cite{batt03}).  The SZ effect is independent of redshift and is
therefore an excellent tracer of large scale structure formation
and evolution. This property has inspired groups to propose blind SZ
surveys and investigate the way they can probe the cosmological
parameters (e.g. \cite{bartlett94}, \cite{barbosa96},
\cite{holder2000}, \cite{da_silva2000}, \cite{kneissl2001},
\cite{xue2001}).  As a consequence, several SZ experiments are either
planned, under construction, or already observing. Some of these
experiments are interferometric arrays like AMIBA \cite[]{lo2001}, AMI
\cite[]{kneissl2001}, SZA~\footnote{//astro.uchicago.edu/sza/} 
\cite[]{mohr2002}. Others are single dish
multi-frequency instruments like Planck
surveyor\footnote{//www.rssd.esa.int/Planck/},
SPT\footnote{//astro.uchicago.edu/spt},
ACT\footnote{//www.hep.upenn.edu/\~{}angelica/act/act.html}, OLIMPO
\cite[]{masi2003}, SUZIE-II
\footnote{//www.stanford.edu/\~{}schurch/suzie\_instrument.html},
ACBAR\footnote{//cosmology.berkeley.edu/group/swlh/acbar/}, MITO 
\footnote{//oberon.roma1.infn.it/mito/} 

The SZ effect can also be used to characterise the galaxy clusters
themselves. \cite{pointecouteau98} proposed  measuring the
intra-cluster gas temperature from the relativistic corrections to the
SZ. This is a particularly important issue in the context of
future SZ surveys for which X-ray counterparts will not be easily
available  and since X-ray temperatures have not been measured for all
clusters. This is because X-ray temperature determination is
expensive, and one cannot expect that all the many clusters to be
observed in future SZ surveys will have X-ray temperature
determination. The SZ effect can also be used to measure cluster
radial peculiar velocities as suggested by \cite{sz80}. This provides
a new distance free measurement of the cluster velocities, which is
very interesting in view of future SZ surveys. However this needs
measurement of the intra-cluster temperature. So far, it is
only for a handful of known rich clusters that attempts have been made
to extract the intra-cluster gas temperature \cite[]{hansen02} or to
extract the peculiar radial velocities
\cite[]{holzapfel97,lamarre98,benson2003}.  It is fair to say that
only upper limits have been obtained, and an actual temperature or
peculiar velocity determination has still to be made.

The SZ effect is a potentially powerful cosmological probe; however,
it is not free of contamination. As a matter of fact, SZ measurements
are contaminated by other astrophysical sources that are mainly of
two types: (i) due to our galaxy (free-free, synchrotron and dust
emission from the milky way) or (ii) due to extra-galactic point
sources (radio and infra-red (IR) galaxies), and even (iii) due to the Cosmic
Microwave background (CMB) itself. The kinetic SZ and primary CMB
fluctuations have the same frequency dependence, and therefore 
will a separation of the
CMB signal and the kinetic effect only be possible through the
relativistic corrections to the kinetic effect.  This would require
both very high sensitivity (of the order $0.1 \mu K$) and additional
observing frequencies at the extrema of the relativistic corrections
to the kinetic effect (near 200 and 500 GHz) and at its cross-over
(near 300 GHz). The effect of the primary CMB anisotropies is well
known (see for example \cite{haehnelt96} and \cite{aghanim97}). It
induces a further uncertainty to the peculiar velocity and limits its
accuracy. This effect intervenes mostly for large beam experiments
like Planck, as small beam experiments are indeed less affected by the
CMB which power is severely damped on a scale of about a
few arcminutes.

These contaminations may be monitored in the context of pointed SZ
observations (i.e. towards known clusters) or follow-up
observations of likely clusters. However, systematic follow-up
observations of all the clusters in the SZ survey will be very 
time-consuming. Ignoring contaminations modifies the effectiveness of a
survey by the loss of certain clusters and the appearance of some
other "fake" artificial clusters. For future surveys accurate
knowledge of the completeness and reliability is thus crucial for
extracting cosmological information. The question of survey
selection function is being studied extensively for the Planck survey
\cite[]{white2003a,geisbusch2004,schafer2004}; however, the systematic
effect from the assumed cluster structures still remains to be studied
carefully \cite[]{birkinshaw2004,hansen04a}.   Contamination of SZ
measurements by extra-galactic sources, especially radio sources, is
not a new problem (see for example reviews by \cite{rephaeli95},
Birkinshaw (1999)). It can be due to radio
emission of the galaxies in the cluster itself
\cite[]{ledlow96,cooray98,lin2002} or to foreground galaxies. More
specifically, the emission of radio sources can dilute the SZ
signal. \cite{holder2002} estimated the expected dilution and its
effect on the SZ power spectrum.  IR dusty galaxies whose emission
dominates at high frequencies may also contaminate SZ measurements. In
particular, gravitational lensing of dusty galaxies causes
enhancement of the confusion noise, which is likely to affect SZ
observations of nearby clusters \cite[]{blain98}.  The effect of
the point sources (both radio and dusty galaxies) have been recently
revisited by \cite{white2003}, who quantified it in terms of an
equivalent noise and computed the associated power spectra. Also
recently, \cite{knox2003} investigated the effects of IR galaxy
contamination on the statistical error-bars of the SZ parameters. 
Contamination of SZ clusters by radio sources at low frequencies
that can be monitored by the interferometric arrays remains potentially
a very important source of errors for single dish experiments. On the
other hand, multi-frequency observations with single dish experiments
should help in solving the problem of contamination.

In this study, we have focused on one category of SZ instruments
planned for SZ surveying, namely the single-dish multi-frequency
experiments. The SZ number counts that will be provided to us by these
surveys will certainly probe and constrain the cosmological
parameters. However, such instruments are theoretically able not only
to measure the SZ effect amplitude  for each detected cluster (through
the Compton parameter) but also its radial peculiar velocity and its
gas temperature. We explored the capabilities of these future SZ
surveys in terms of measuring the three cluster SZ parameters: Compton
parameter, intra-cluster gas temperature, and radial peculiar velocity
independently of other observations.  In a previous work
\cite[]{aghanim03}, we investigated the effects of the cluster
parameter degeneracies on the of the above-mentioned
cluster SZ parameters in terms of error-bars.  Here we were interested in
the effects of the different sources of contamination on the
measurement of these parameters. The contamination did not affect the
statistical error-bars strongly. 

The main question we are seeking to
answer in this article is: {\it How big are the systematic errors
due to non-removable contamination?} We show that the systematic
errors are very important. They are by far the dominant source of
trouble for the future single-dish SZ blind surveys when these surveys
are used alone to estimate cluster parameters of unresolved clusters.
We start by presenting our models for the major astrophysical
contaminants in Sect. 2, and we then discuss the SZ parameter extraction
technique and the sample of galaxy clusters used in our study in
Sect. 3.  In Sects. 4 and 5, we present the results for the three SZ
experiments under consideration in our study, and discuss the results
in Sect. 6.  We finally offer our conclusions in Sect. 7.

\section{Modeling the contaminants}
The contaminations strongly depend on the observing frequency. For
galactic foregrounds
the frequency dependance is given by the spectral energy distribution on
the contaminants.
For extra-galactic sources the frequency dependance results from the
combination of
the redshift evolution of the luminosity function with the spectral
energy distribution of the sources.
Moreover, the contamination levels, especially of the
galactic foregrounds, also depend on the observed region of the sky
and on the spatial distribution of the signal (e.g. the power spectrum
of dust emission).  Multi-frequency observations should help in
reducing the contamination level through component separation, but we
expect some 
contamination of the measurements to remain.  This is
rendered even more critical due to the rather large uncertainties in
the emission models used to describe the contaminating sources.

In order to address the question of statistical and systematic errors
on the cluster SZ parameter extraction due to non-removable
contaminations, we computed the level of contamination at each source
for a set of pre-defined frequencies.  We based our study on an
ensemble of planned SZ instruments which thus define our choice of
frequencies.  We distinguished between a large beam ($\sim 5$
arcmin at best) experiment like Planck, and small beam ($\sim2$
arcmin) experiments ACT-like. For the Planck instrument, we excluded
from our analysis the 857 GHz channel which is totally dominated by IR
emission from dust in the inter-stellar medium (ISM) and in high
redshift galaxies.  The ACT-like experiment had three observing
frequencies 145, 225, and 265 GHz. We assumed an optimistic
sensitivity of 2 $\rm \mu$K for each channel and a beam of 1.7 arcmin
at 145 GHz,  along with beam-size scaling as 1/frequency.

\subsection{Dust in the ISM}

Thermal emission from dust particles in the ISM is an important source
of radiation in the IR and submm domains ($\nu\ge100$ GHz). It thus
contaminates the measurement of the SZ effect at high frequencies. We
therefore did not compute its contribution to the 30 and 44~GHz
channels of Planck. Dust emission is usually modeled by a modified
blackbody spectrum: $S_\nu^\mathrm{dust}=\tau \times B_\nu(T_{\rm d})$
where $B_\nu(T_{\rm d})$ is the blackbody spectrum, $T_{\rm d}$ the
dust temperature, and $\tau$ is the emissivity. The emissivity is
proportional $\nu^\beta$ where $\beta$ is the spectral index.  The
values for $\beta$ and $T_{\rm d}$ are not unique but vary from dense
clouds to diffuse gas. In the diffuse part of the sky
($N_\mathrm{HI}$$\le$5 10$^{20}$ atoms/cm$^{-2}$) the dust emission
spectrum from the far-IR to the submillimeter does not spatially
vary \cite[]{boulanger96,finkbeiner99,lagache2003}.  Because of the
difficulty of modeling dust emission properly, we instead computed the
observed dust emission spectrum as in \cite{lagache2003} from a
combination of COBE and WMAP measurements.  We derived the mean
spectrum of the dust emission correlated with HI in the diffuse part
of the sky for $|b|>30^{\rm o}$. This spectrum, which represents the
far-IR to mm emission for $N_\mathrm{HI}$=10$^{20}$ atoms/cm$^{2}$, was
used as a reference. The colors, i.e. the shape, of this spectrum are
in very good agreement with \cite{finkbeiner99}.\\

For the dust emission spatial distribution, our analysis relied on the power
spectrum measured in HI \cite[]{miville2002} and in
far-IR \cite[]{gautier92,miville2004}, 
which is a steep power law power spectrum with no break.  We thus
used the same power spectrum $\propto k^{-3}$ that describes the dust
distribution over a very large range of angular scales.\\

To evaluate the  fluctuation level of dust emission in the case of
SZ experiments, we computed the dust emission power spectrum (in Jy$^2$/sr)
at each observing frequency using the following expression:
$$
P(k) = 4.8\,10^5\times(N_\mathrm{HI}/10^{20})^{2.1}\times(k/0.01)^{-3}
\times(B_\lambda/B_{100})^2
$$ where $k$ is in arcmin$^{-1}$, $N_\mathrm{HI}$ the HI column
density in atoms/cm$^{-2}$, and the $B_\lambda/B_{100}$ the color
given at each observing wavelength $\lambda=h\nu/c$ (in $\mu$m). In
our calculations, we used the column density associated with the
cleanest 40\% of the sky that corresponds to $N_\mathrm{HI} \sim$
2.31 10$^{20}$ atoms/cm$^{2}$.  We finally computed the fluctuations
of the dust emission in the beam of the considered experiments which
are given by $\sigma^2 = \int P(k)2\pi k\,dk$ Jy$^2$/sr$^2$. For a
$k^{-3}$ power spectrum, the previous integral was dominated by the
lower limit of $k$, which is taken to be equal to 1/FWHM.

The dust fluctuations cannot be removed and perfectly corrected for.
In Tables \ref{tab:plck} and \ref{tab:act} we show 
the level of contamination from dust
fluctuations for Planck and for the ACT-like experiment
for two cases: (i)  where 30\% of the
fluctuations (i.e. 0.3$\sigma$) remain unremoved, and (ii) where
100\% remains. This 30\% is the realistic level of contamination 
when independent tracers of dust emission (such as HI) are used to remove
the dust contribution.
The latter case (100\%) is displayed to show how the
maximum contribution affects the results. In both cases (30 and 100\%
fluctuations), the contamination is computed for the 40\% cleanest
region of the sky in terms of gas column density, as found outside
the galactic plane and in bright molecular clouds. 

\begin{table*}
\begin{center}
\begin{tabular}{|c||c|c|c|c|c|c|c|c|}
\hline
 Frequency (GHz)& 30& 44& 70&100&143&217&353&545 \\
\hline
\hline
30\% Dust Fluctuations & -- & -- &3.96&13.09&33.76&144.69&704.14&2218.54\\
100\% Dust Fluctuations & -- & -- &13.22&43.65&112.53&482.29&2347.15&7395.14\\
\hline
\hline
Poisson CIB fluctuations & -- & -- &311.27&583.41&1445.88&4426.90&15279.8&39082.8\\
Poisson+Correlated CIB fluctuations & -- & -- & 696.02&1304.54&3233.09&9898.85&34166.7&87391.8\\
\hline
\hline
Radio source fluctuations & 1246.7& 1729.3& 2830.4& 3210.0& 3204.5& 3598.3& -- & -- \\
\hline
\end{tabular}
\end{center}
\caption{Level of contaminations (in Jy/sr) for Planck. 
The first two lines are for the dust emission fluctuations in the 40\%
cleanest region of the sky, assuming a $k^{-3}$ spectrum when 30\% and
100\% of the fluctuations remain unremoved. The second two lines are for
the CIB fluctuations. We computed the case where the IR galaxies
are Poisson distributed and the case where we also  take the
correlations between sources into account. In this case, $\sigma_{\rm
tot}^{\rm CIB}=\sqrt{5 \times \sigma_{\rm Poisson}^2}$. The last line is for
the fluctuations due to the unresolved radio, while resolved radio sources
represent 12\% to 15\% of the background.}
\label{tab:plck}
\end{table*}

\subsection{Infra-red or ``dusty'' galaxies}

Star-forming galaxies not only emit in the near-IR, optical, and UV
domains, but also in the far-IR and submm. This emission is associated
with the radiation that is absorbed and re-emitted by the dust in
galaxies. For local galaxies, up to one third of the output emission
lies in the far-IR domain. Star-forming galaxies can have much larger
fractions of their emitted radiation in the far-IR (up to 90\%).  These
galaxies are called ``IR galaxies'' in the following.  The total
emission from unresolved and/or faint IR galaxies is responsible for
the Cosmic Infra-red Background (CIB) (Puget et al. 1996), which
contains as much power as the optical and UV backgrounds. The
fluctuations of the CIB that are now detected
\cite[]{lagache2000a,matsuhara2000,miville2002,kiss2001} are
one of the major contaminants of the SZ measurements at high
frequencies ($\ge 100$ GHz). As in the dust case, we will not
compute the contribution at the 30 and 44~GHz channels of Planck.

In order to evaluate the level of contamination by fluctuations of
the CIB, i.e. unresolved dusty galaxies, we used the model of
\cite{lagache2003a,lagache2004}, which provided us with the
number counts of IR galaxies at each frequency of the SZ experiments
(Planck and ACT-like). We first estimated the IR background emission
from the dusty galaxies by integrating the number counts below a given
detection limit $S^\mathrm{IR}_\mathrm{lim}$. We then computed the {\it
rms} fluctuations of the CIB due to unresolved sources in the beam of
the SZ instruments.  For this step, we considered two cases, one where
the sources follow a Poisson distribution and a second where the
correlation between sources is added to the shot
noise. \cite{song2003} have investigated the effect of correlations at
high frequencies, and find it boosts the CIB shot noise
fluctuations by a factor of 1.7 on average.  Due to the lack of
observational constraints or precise models for the correlations for
all frequencies, we simply assumed that the correlated part of
the CIB fluctuations is twice as large as the Poisson part for all
frequencies and for all the experiments.\footnote{We have taken this
factor 2 from the modeling of the clustering of the CIB anisotropies
of \cite{knox2001}.}  It is worth noting that we did not take 
possible enhancement by lensing into
account. Our estimate of the CIB
contribution is thus likely to be slightly underestimated especially
for low redshift clusters.

We estimated the contribution of the CIB fluctuations to the SZ
measurements in two cases: one case where only 1\% of the dusty
galaxies are resolved, which is the case predicted for Planck surveyor
\cite[]{lagache2003a}. In the second case for the  ACT-like experiment,
we arbitrarily assumed 50\% of the CIB is
in resolved IR galaxies. The level of fluctuations are given for
Planck and ACT-like in Tables \ref{tab:plck} and
\ref{tab:act}, respectively.

\subsection{Radio sources}\label{sec:radio}
The radio emission of radio sources is dominated by synchrotron
emission from relativistic electrons spiraling around magnetic fields
in the sources. The synchrotron spectrum follows a power law
$\nu^\alpha$, with $-1<\alpha<-0.5$. However, other types
of sources exist such as those with ``inverted spectra'' (with $\alpha>0$).
In general, the spectra are not simple power laws with fixed indexes
\cite[]{herbig92}, but have curved spectra instead; moreover some of
the radio sources may be variable. It is therefore very hard to
extrapolate their fluxes from one frequency to another. Additionally,
radio sources are well catalogued at low frequencies thanks to NVSS
\cite[]{condon98} and FIRST \cite[]{white97}, but there are no all-sky
surveys of extra-galactic sources at frequencies $>5$ GHz apart from
the WMAP observations at 41 GHz \cite[]{bennett2003}. As a result,
properties of the radio sources at high frequencies are poorly
known. In the context of CMB observations, several studies have aimed
at predicting the radio source counts at frequencies higher than 5 GHz
from known sources \cite[]{toffolatti98,sokasian2001}.  These studies
allow us, in particular, to assess the level of contamination due to radio
sources at the frequencies of interest for the CMB, where they are the
dominant contamination between 30 and 217 GHz. However, one has
to keep in mind that extrapolations of radio spectra from 5 GHz to high
frequencies is quite a difficult task.

In our study, we compute the level of contamination of unresolved
radio sources per frequency by integrating the source counts below a
flux limit, $S^\mathrm{radio}_\mathrm{lim}$, determined for each
frequency.  Assuming Poisson distribution for the sources, we then
evaluate the {\it rms} fluctuations due to unresolved radio sources in
the beam of each SZ experiment (Planck and ACT-like). We did not take
 the effect of the correlation of radio sources into account since it
is rather weak. It is worth noting in the case of Planck, in
particular, that the source fluctuations increase with frequency
although radio sources are getting fainter. This is due to the
decrease of the beam-size with frequency from 33 arcmin at 30 GHz to
5.5 arcmin at 217 GHz.  In practice, we use the radio-source counts
that were kindly provided by L. Toffolatti and are based on the model
of \cite{toffolatti98}. This model was found to be in quite good
agreement with  recent observations of radio sources by WMAP at 41
GHz. It is thus likely to provide us with a good prediction of
the radio counts at higher frequencies. As mentioned above, to
estimate the level of contamination from unresolved sources we need
the flux limit at each frequency for the detection of radio
sources. In the case of Planck frequencies, \cite{vielva2001}
provided us with the flux limits $S^\mathrm{radio}_\mathrm{lim}$ above which
the radio sources are detected. They were computed using simulated observed
sky taking  all galactic and extra-galactic emissions into account,
together with instrumental noise. These flux limits correspond to
$5\sigma$ detections. As a consequence, the residual fluctuations
computed in Table \ref{tab:plck} are likely to be overestimated. 

Using the flux limits of \cite{vielva2001} ranging from
$S^\mathrm{radio}_\mathrm{lim}=540$ mJy (at 30 GHz) to
$S^\mathrm{radio}_\mathrm{lim}=240$ mJy (at 217 GHz), we found
that about 12\% to 15\% of the radio background was resolved into
individual sources (detected at $5\sigma$) by Planck.  The levels of
contamination are summarised in Table \ref{tab:plck}.  For ACT-like,
we investigated the effect of unresolved radio sources on the cluster
parameter measurement by assuming arbitrarily and optimistically that
50\% of the radio background is resolved into sources.  The numbers
are given in Table \ref{tab:act}.  Contrary to what is stated in
\cite{knox2003}, the radio source contamination
always dominates over the fluctuations from IR galaxies up to 217 GHz
when the latter are Poisson distributed, and up to 143 GHz when we
take  the correlations of IR galaxies into account. We did not compute
the contributions at 353 and 545~GHz for Planck since they become
quite subdominant.

\subsection{Synchrotron and Free-Free}

The synchrotron emission in our galaxy is due to the emission of free
electrons spiraling around the lines of magnetic field.  This emission
is difficult to model because of  poor knowledge  of the
galactic magnetic field itself.  Synchrotron emission is expected to
dominate at low frequencies ($< 80$ GHz). \cite{bennett2003} argue
that this emission in fact dominates over all other CMB foregrounds in
the WMAP frequency bands. But at high galactic latitudes in the
diffuse sky (where the SZ observations will be conducted),
\cite{lagache2003} show that the WMAP emission is dominated by the
so-called ``anomalous microwave emission'' associated with the small
interstellar transiently heated dust grains. This component was
included in our measured dust spectrum.

We did not include the
free-free emission that is, in the diffuse sky at high latitudes,
much smaller than the amplitude of the anomalous microwave component.
We also ignore a possible contamination from ultra relativistic
electrons~\citep{ensslin1}, which is at best marginally detectable
with future SZ surveys~\citep{ensslin2}.
\begin{table*}
\begin{center}
\begin{tabular}{|c|c|c|c|}
\hline
 Frequency (GHz)& 145&225&265 \\
\hline
\hline
30\% Dust Fluctuations & 19.65& 67.44& 128.62\\
100\% Dust Fluctuations & 89.17&224.79 &428.74\\
\hline
\hline
Poisson CIB Fluctuations & 1225.19& 5430.84& 10877.9\\
Poisson+Correlated CIB Fluctuations & 2739.61 &12143.7 &24323.7\\
\hline
\hline
Radio sources Fluctuations & 2265.42&2873.19&2821.53\\
\hline
\end{tabular}
\end{center}
\caption{Level of contaminations in Jy/sr for the ACT experiment.  The
first two lines are for the dust emission fluctuations  in the
40\% cleanest region of the sky when 30\% and 100\% of the
fluctuations remain unremoved. The second two lines are for
fluctuations of the 50\% unresolved IR galaxies, when they are Poisson
distributed and where we also take  the correlations into account. In
this case, $\sigma_{\rm tot}^{\rm CIB}=\sqrt{5 \times \sigma_{\rm
Poisson}^2}$. The last line represents the fluctuations due to
unresolved radio sources when 50\% of the background is resolved.}
\label{tab:act}
\end{table*}

\section{SZ parameters\label{sec:sz}}

The SZ effect is traditionally separated into two components according
to the origin of the scattering of the same electrons
\begin{eqnarray}
\frac{\Delta I(x)}{I_0} &=& \Delta I_{{\rm thermal}} \, (x, y, T_e)
+ \Delta I_{{\rm kinetic}} \, (x, \tau, v_p)
\label{eq:deltai}
\\ 
&=& y \, \left( g(x)+\delta_T(x,T_e) \right) 
-  \beta \tau \, 
h(x)
\,, \nonumber
\end{eqnarray}
with $x = h \nu/kT_{\rm cmb}$ and $I_0 = 2 (kT_{\rm cmb})^3/(hc)^2$ where
$T_{\rm cmb} = 2.725$ K.  The first term on the rhs of
Eq.~(\ref{eq:deltai}) is the thermal distortion~\citep{sz72}
with the
non-relativistic spectral shape
\begin{equation}
g(x) = \frac{x^4 \,e^x}{(e^x -1)^2} \left( x\,\frac{e^x + 1}{e^x -1} -
4 \right)\,,
\end{equation}
and the magnitude is given by the Compton parameter 
\begin{equation}
\label{eq:ygas}
y = \frac{\sigma_{\rm T}}{m_e \, c^2}\, \int\!\! {\rm d}l\, n_{e}
\,kT_e \, ,
\end{equation}
where $m_e$ and $n_e$ are masses and number density of the electrons,
$\sigma_{\rm T}$ is the Thomson cross section, and the integral is along the
line of sight.  For non-relativistic electrons one has
$\delta_T(x,T_e) = 0$, but for hot clusters the relativistic electrons
will slightly modify the thermal SZ effect~\citep{wright79}.
The spectral shape of $\delta _T(x,T_e)$ is easily calculated
numerically~\citep{rep95,dolgov01,shimon03,sandoval03,itohnozawa03}.
For the Planck frequencies this is easily implemented using the simple
fits presented in \cite{diego03}.

The kinetic distortions~\citep{sz80} in the last term of Eq.~(\ref{eq:deltai}) 
have the spectral shape
\begin{equation}
h(x) = \frac{x^4 \,e^x}{(e^x -1)^2} \, ,
\end{equation}
and the magnitude depends on $\beta = v_p/c$, the average
line-of-sight streaming velocity of the thermal gas (positive if the
gas is approaching the observer), and the Thomson optical depth
\begin{equation}
\label{eq:barbeta}
\tau = \sigma_{\rm T} \int\!\! {\rm d}l\, n_{e} \, .
\end{equation}
The minor thermal corrections to the kinetic effect 
are negligible with existing sensitivity.
Thus, when the intra-cluster gas can be assumed isothermal one has $y
= \tau kT_e/(m_ec^2)$, while for non-isothermal clusters things are slightly
more complicated~\citep{hansen04}.

Given the different spectral signatures of $g(x), h(x)$, and $\delta_T
(x,T_e)$, it is straightforward to separate the physical variables
$y,v_p$, and $T_e$ from sensitive multi-frequency observations.  This
separation is easily understood as follows.  To a good approximation
the magnitude of the SZ effect gives the $y$ parameter, which is
measured at the maximum of $g(x)$ (near 129 or 370 GHz). The velocity
times optical depth, $\beta \tau$, is measured at the zero-point of
$g(x)$ (near 217~GHz). Finally,  electron temperature $T_e$ is
measured using the zero point and maxima of $\delta_T$ (near 190, 330,
475 GHz).  Hence, under the assumption of isothermality, these
observables can be combined to provide the physical parameters $y,
v_p$, and $T_e$.  It is thus clear that with at least 3 well-placed
observing frequencies one can separate all 3 physical variables by
using only the SZ observations. The first marginal determination of all 3
SZ parameters was made for the hot cluster A2163~\citep{hansen02}
using data from SUZIE~\citep{holzapfel97} and
BIMA/OVRO~\citep{laroque02}.  However, as was shown in
\cite{aghanim03} additional observing frequencies are needed to break
the intrinsic degeneracies between the SZ cluster parameters.

\subsection{The practical parameter extraction}
In order to find the SZ parameters for a given cluster we modified the
publicly available SZ-parameter extraction code {\tt SASZ}.  For a
given ``SZ observation'' this code fast and accurately determines the
central value for the 3 SZ parameters ($y, v_p$, and $T_e$), and
corresponding $1\sigma$ error-bars~\citep{hansen03}.  These error-bars
reflect solely the statistical error-bars due to the observational
sensitivity (which for Planck or ACT-like is on the order of a few
$\mu K$ for each channel).  This parameter extraction code is based on
the stochastic method of simulated annealing, and we chose the
cooling scheme of {\tt SASZ} in such a way that the central values of
the 3 SZ parameters of all the clusters in our sample were determined
to better than $0.1\%$ accuracy.

\subsection{The galaxy cluster sample}
To finally realistically answer the question of how big the systematic 
errors due to non-removable contamination are, one must first
of all consider a
large sample of clusters. We simulated clusters in bins of mass and
redshift, using an extended Press-Schechter formalism that provides
realistic temperatures, Compton parameters, and peculiar velocities
(for details see~\cite{aghanim01} and \cite{aghanim03}).  In our
study, we considered clusters with redshifts between 0.2 and
3. Within this range of redshifts, not all clusters fill the beam of
the instrument (i.e. $10R_c=$beam-size, with $R_c$ the cluster core
radius), which sets a limit to the cluster mass range. We illustrate
this point in the case of Planck, for which we set a conservative value
of 9 arcmin for the effective beam-size of the instrument for cluster
observations. In this case, all clusters with total masses between
$5.\,10^{13}$ and $10^{16}$ solar masses and redshifts $z > 0.5$ fill
the beam. However, at lower redshifts $z=0.2$(0.3 and 0.4) only
clusters with masses above $1.4\,10^{15}$($4.\,10^{15}$ and
$8.\,10^{15}$) solar masses fulfill this condition.  Specifically we
consider 500 hot clusters with temperatures in the range $3.5 {\rm
keV} < T_e < 7.5 {\rm keV}$, central Compton parameter $-4.7 <
log(y) < -3.7$, and peculiar velocity $ -1600 {\rm km/sec} < v_p <
1600 {\rm km/sec}$. For each cluster we thus have the true values,
$y_{\rm true}, T_{\rm true}$, and $v_{\rm true}$, from which one can
calculate the true SZ signal, $\Delta I_{\rm SZ}$.

Under the assumption that all contaminations have been removed
completely, the total signal is equal to the SZ signal (ignoring the
noise), $\Delta I_{\rm total} = \Delta I_{\rm SZ}$. Thereby, one can
import the total signal to {\tt SASZ}, which then finds the best fit
parameters, $y_{\rm derived}, T_{\rm derived}$, and $v_{\rm derived}$.
For all the 500 clusters in our sample the central values of the
parameters have been  determined to better than $0.1\%$ accuracy.

Now, the interesting question is which central values are found for
the SZ parameters when the contaminations have not been removed
completely. For instance, if the dust has not been removed completely,
then there will be an additional signal, $\Delta I_{\rm dust}$. In
that way the total signal is $\Delta I_{\rm total} = \Delta I_{\rm SZ}
+ \Delta I_{\rm dust}$.  Believing this signal is purely from SZ will
induce a systematic error, $y_{\rm derived} = y_{\rm true} + \delta
y_{\rm systematic}$, but it will not  significantly 
change the statistical error due
to instrumental sensitivity. We explicitly compared the
statistical error-bars due to the instrumental sensitivity for
slightly different cluster parameters, and found the effect on the
statistical error-bars very moderate (a few percent).  We will now
address the question of how big these systematic errors are.
We show that these systematic errors for future planned SZ
surveys can be significantly larger than the statistical errors due to
instrumental sensitivity.

\section{Planck surveyor case}

As the first example we consider the all-sky  Planck survey.  The
expected error-bars on the SZ parameters from Planck observations
without taking  the contaminations into account have already been
studied \cite[]{aghanim03}. Due to its wide frequency range, earlier
predictions estimated that the Compton parameter should be easily
measured and that the cluster peculiar velocity should be very
marginally estimated with an error of about 500-700 km/s.
In particular the importance of non-Gaussian error-bars was discussed
in detail in \cite{aghanim03}.  In our study, we considered
a simplification to the real error-bars.  Instead of having different
upper and lower error-bars, we  simply took the largest of the
two; that is, for all observables we  consider  $T ^{+\delta T_{\rm
up}}_{-\delta T_{\rm low}}$, with $\delta T = \delta T_{\rm up} =
\delta T_{\rm low} = {\rm max}(\delta T_{\rm up},\delta T_{\rm low})$.

\begin{figure}[htb]
\begin{center}
\psfig{figure=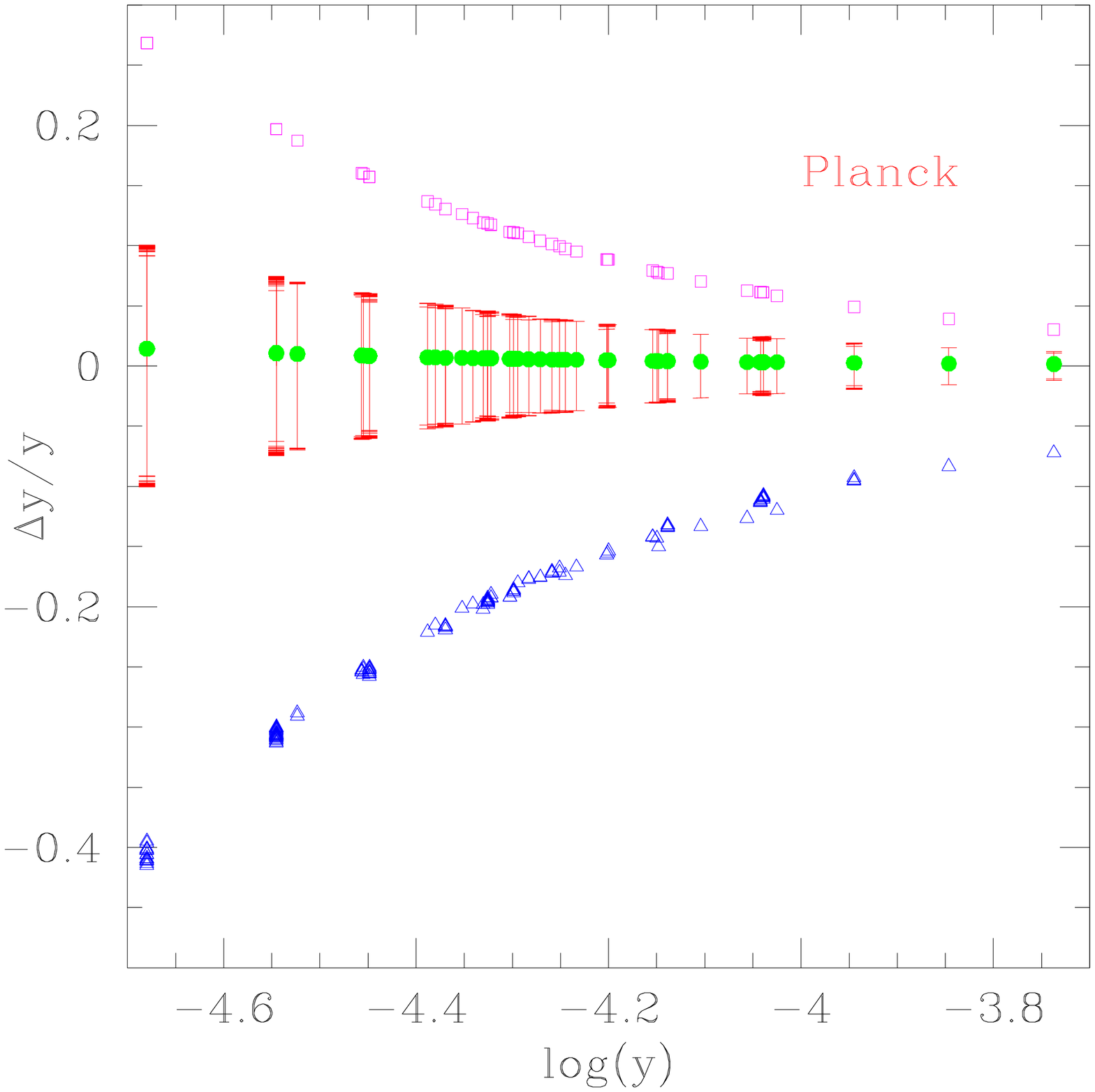,width=0.8 \textwidth} 
\end{center}
\vspace{-0.8cm}
\caption{The expected error-bars (red) on the Compton parameter for
the Planck experiment, without contamination.  The open triangles
(blue) show the effect due to non-removed radio
contamination. The open squares (purple) are for correlated IR
galaxies, and the filled circles (green) are when none of the dust
emission fluctuations have been removed. }
\label{fig:fig.con.y} 
\end{figure}

In Fig.~\ref{fig:fig.con.y} we present  the expected Planck statistical
error-bars on $y$ within this simplification.  In the idealised case
where no contamination is taken into account,  Compton parameter
$y$ is found to an accuracy of $1\%$ for the brightest clusters with
$y = 10^{-3.7}$ and to $10\%$ for the dimmer clusters in our sample
with $y = 10^{-4.7}$.  The temperature, it is determined at best
with a few keV, and more typically with $5-10$ keV, statistical
error-bars.  The size of the statistical error-bars on $v_p$ depends
very much on the actual value of $v_p$, and goes from few hundred km/s,
for the best case, to 2000-3000 km/s for the worst. It is worth
noting that for our implementation of the extended Press-Schechter
formalism (Sect. 3.2) we binned the data in cluster mass and
redshift. This results in a quantization of the values (e.g. log($y$))
and gives the impression of a binning in the figures.  This has no
impact on the results, since all we need is a sample of cluster
parameters that roughly span the ranges expected in a standard
cosmology.

We next added the various non-removed contaminations to the signal to
investigate their effects on the measured parameters.  More
explicitly, with a pure SZ signal, $\Delta I_\mathrm{measured}=\Delta
I_\mathrm{sz}$, one would derive the ``true'' SZ parameters, whereas
with a contaminated signal, $\Delta I_\mathrm{obs} = \Delta
I_\mathrm{sz} + \Delta I_\mathrm{contam}$ one  finds incorrect
derived SZ parameters. Namely, the contamination causes systematic
shifts. In order to quantify the importance of such systematic shifts,
we compared it to the statistical error-bars, given above. We 
compared the value of e.g. $T_\mathrm{derived} - T_\mathrm{true}$ to
$\delta T$. Since the statistical error-bars from above are
conservative, then our findings represent lower limits to the
magnitude of the systematic shift on the SZ parameters due to
contamination. Our sample contains 500 clusters, and we considered
(for $T_e$, for example) the {\it rms} value, over the sample, of the
ratio $T_\mathrm{derived} - T_\mathrm{true}$ to $\delta T$, which is
given by:
\begin{equation}
 \epsilon _T =\frac{1}{N^{1/2}}\left( \sum^N_{i=1} \left(
\frac{T_i^\mathrm{derived}-T_i^{\rm true}}{\delta T_i}\right)^2
\right)^{1/2}\,.
\label{eq:mistake}
\end{equation}

\begin{figure}[htb]
\begin{center}
\psfig{figure=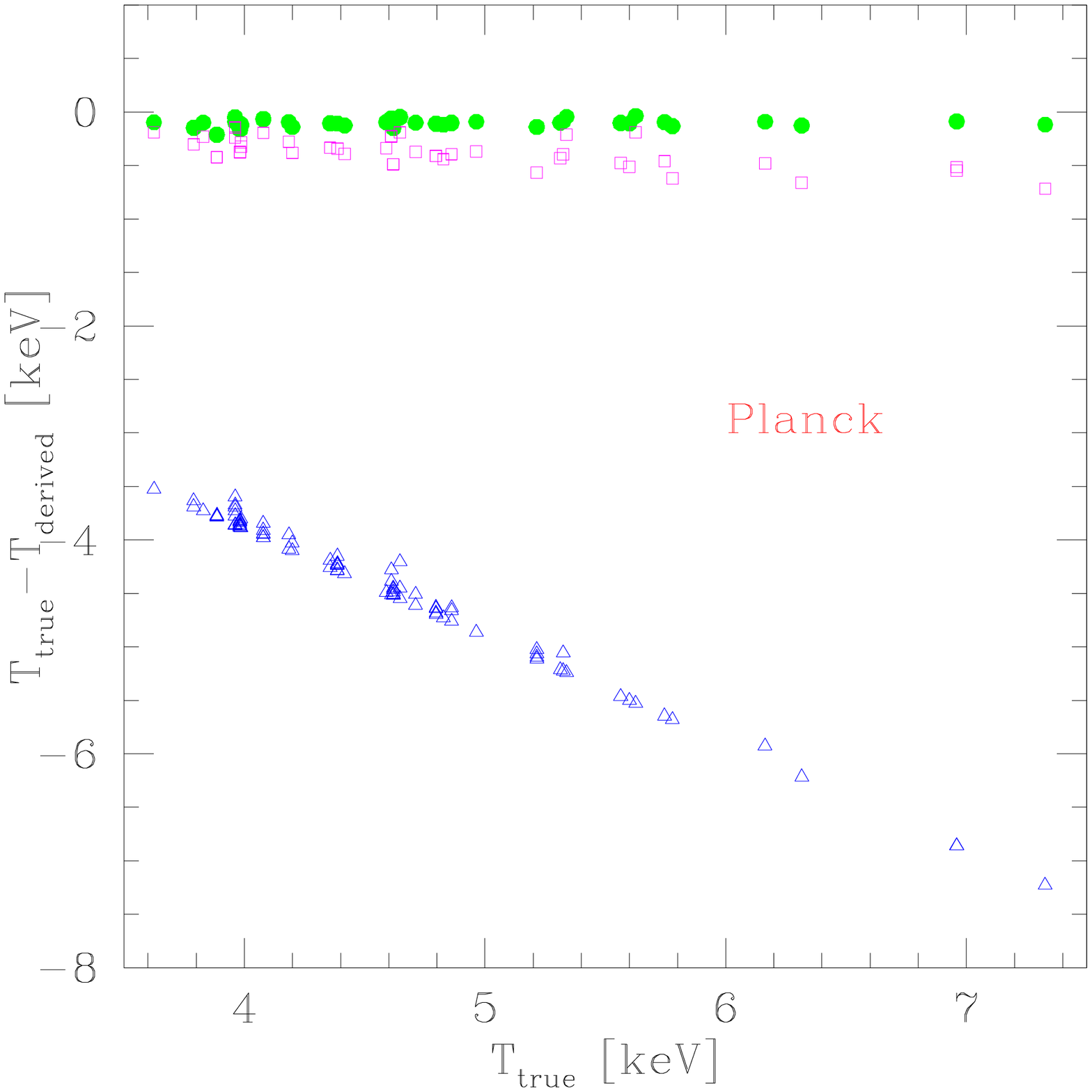,width=0.8 \textwidth} 
\end{center}
\vspace{-0.8cm}
\caption{Derived electron temperature in keV.  Open triangles (blue)
are for radio contamination, open squares (purple) are for
Poisson+correlated IR galaxies, and filled circles (green) are for dust
fluctuations. Statistical error-bars (of the order few to ten keV) are
not shown in order to avoid confusion.}
\label{fig:fig.con.t}
\end{figure}

\begin{figure}[htb]
\begin{center}
\psfig{figure=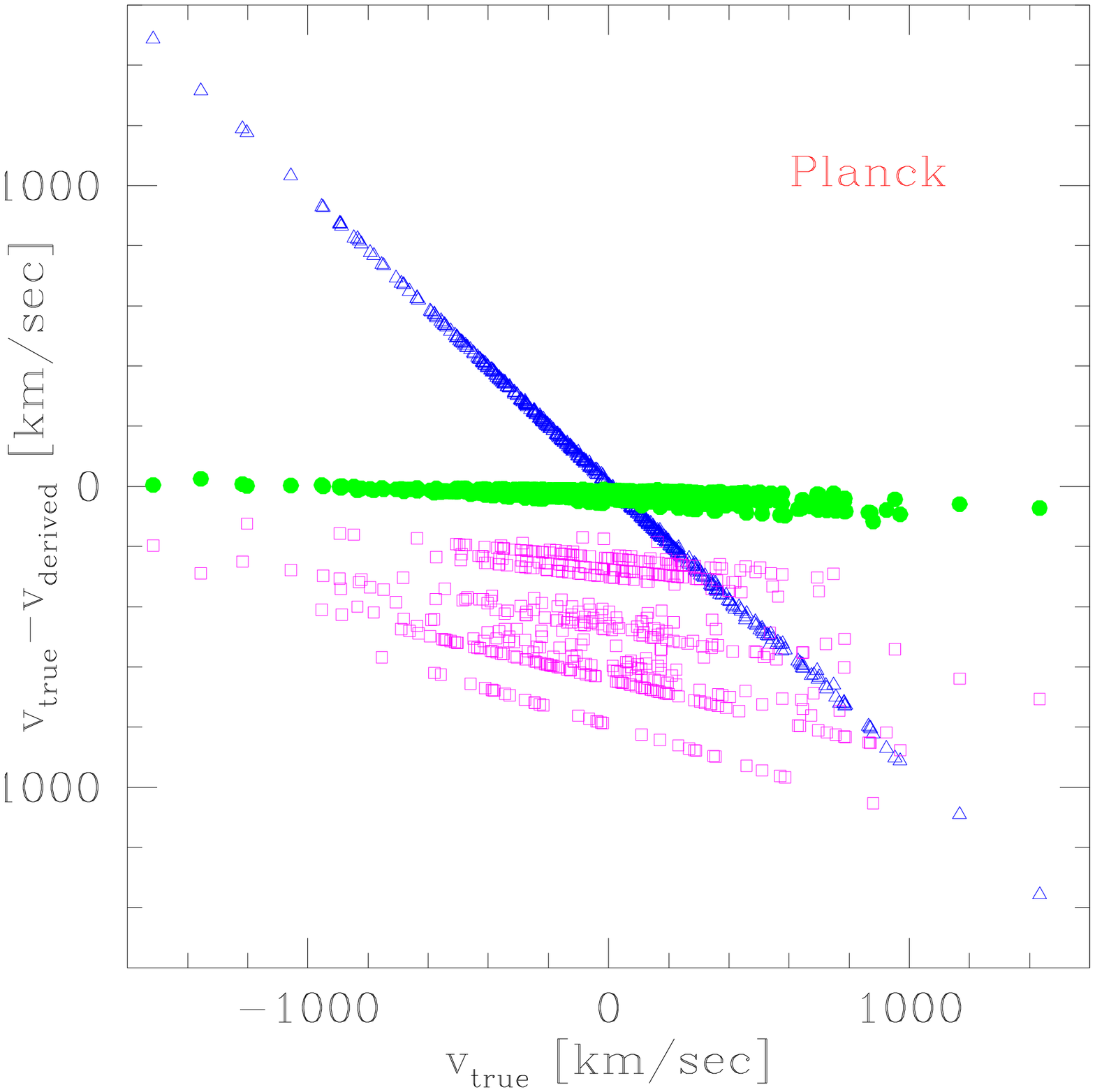,width=0.8 \textwidth} 
\end{center}
\vspace{-0.8cm}
\caption{Upper limit to the expected effect of contamination in
the derived peculiar velocity in km/s.  Open triangles (blue) are for
radio contamination, open squares (purple)  for Poisson+correlated IR
galaxies, and filled circles (green) for dust fluctuation.
Statistical error-bars (on  the order of a few 100 to a few 1000 km/sec) are
not shown, in order to avoid confusion.}
\label{fig:fig.con.vp} 
\end{figure}

We present the results in
Figs.~\ref{fig:fig.con.y}-\ref{fig:fig.con.vp} where we plot the
central values of the derived parameters when adding the
contaminations. In the case of interstellar dust on the $40 \%$
cleanest part of the sky (filled circles), the systematic shifts are
found to be quite small when $70\%$ of the fluctuations are
removed. The effect is small even when no dust fluctuations are
removed. We found $\epsilon_y(\mathrm{dust}) \approx
\epsilon_v(\mathrm{dust}) \approx 0.15$, and the temperature is only
very marginally affected.  We can therefore conclude that the
contamination from the fluctuations of interstellar dust emission will
not be a major problem for estimating the three SZ parameters ($y$,
$T_e$, and $v_p$) in the Planck case.

The situation is completely different when we considered the
contamination due to the fluctuations of the IR background (CIB). In
that case, we found $\epsilon_y(\mathrm{CIB}) \approx 2.5$,
$\epsilon_v(\mathrm{CIB}) \sim 2$, and found minor temperature
shift.  We  also checked the effect of contamination by CIB
Poisson fluctuations only. We found that the {\it rms} ratio between
systematic and statistical errors are about $0.5$ for $y$ and $v_p$ and
negligible for the temperature.

The situation is even worse when we consider contamination from
fluctuations due to the unresolved and unremoved radio sources.  In
this case, we found a very large contamination of the Compton parameter
$\epsilon_y(\mathrm{radio}) \approx 5$, $\epsilon_T(\mathrm{radio})
\sim \epsilon_v(\mathrm{radio}) \sim 1$. As mentioned above, the
level of radio fluctuations is likely to be overestimated since it was
computed using flux limits for $5\sigma$ detections. We redid our
analysis, assuming 20\% of the radio sources to be resolved (less
conservative than the 12-15\% of Vielva et al. (2001)), in which case
the Compton parameter are less affected by the radio contamination,
$\epsilon_y(\mathrm{radio}) \approx 3$. When dust, CIB, and radio
fluctuations are added there is a slight improvement of the
determination of the Compton parameter. We find $\epsilon_y=1.7$;
however, this does not improve determination of the temperature
and velocities. We emphasize that this is an accidental partial
cancellation, which cannot be counted on in general.

\begin{figure}[htb]
\begin{center}
\psfig{figure=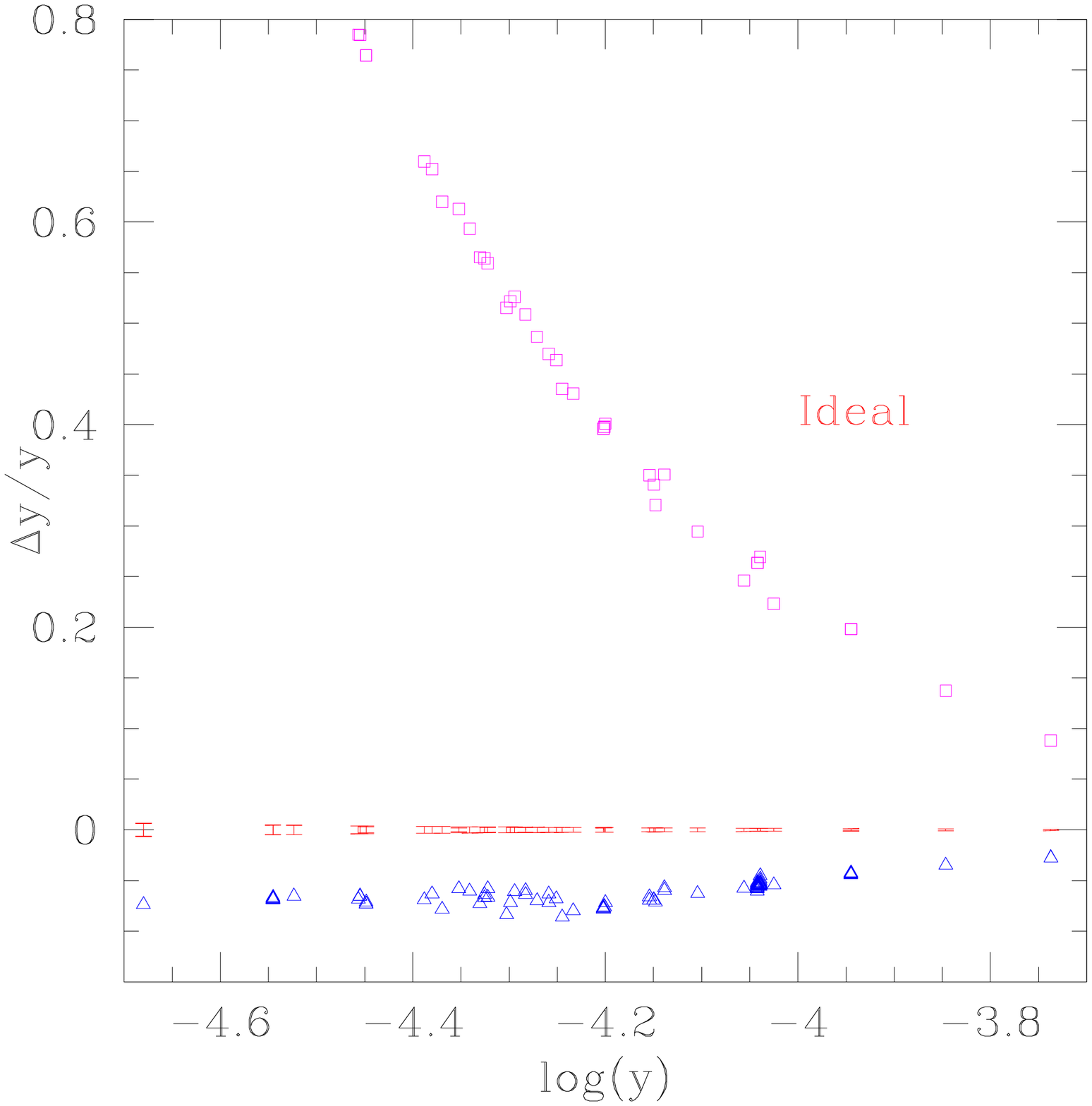,width=0.8 \textwidth} 
\end{center}
\vspace{-0.8cm}
\caption{Compton parameter for an ideal experiment with 4 observing
frequencies at 100, 150, 217, and 270 GHz, and with $0.1\mu$K
sensitivity.  Open triangles (blue)  for radio contamination; open
squares (purple) are for Poisson+correlated IR galaxies.  The very small
statistical error-bars without contamination are shown in red.
Systematic shift from dust fluctuations is very small and is not shown
as it would obscure the statistical error-bars.}
\label{fig:SPT.con.y}
\end{figure}

\begin{figure}[htb]
\begin{center}
\psfig{figure=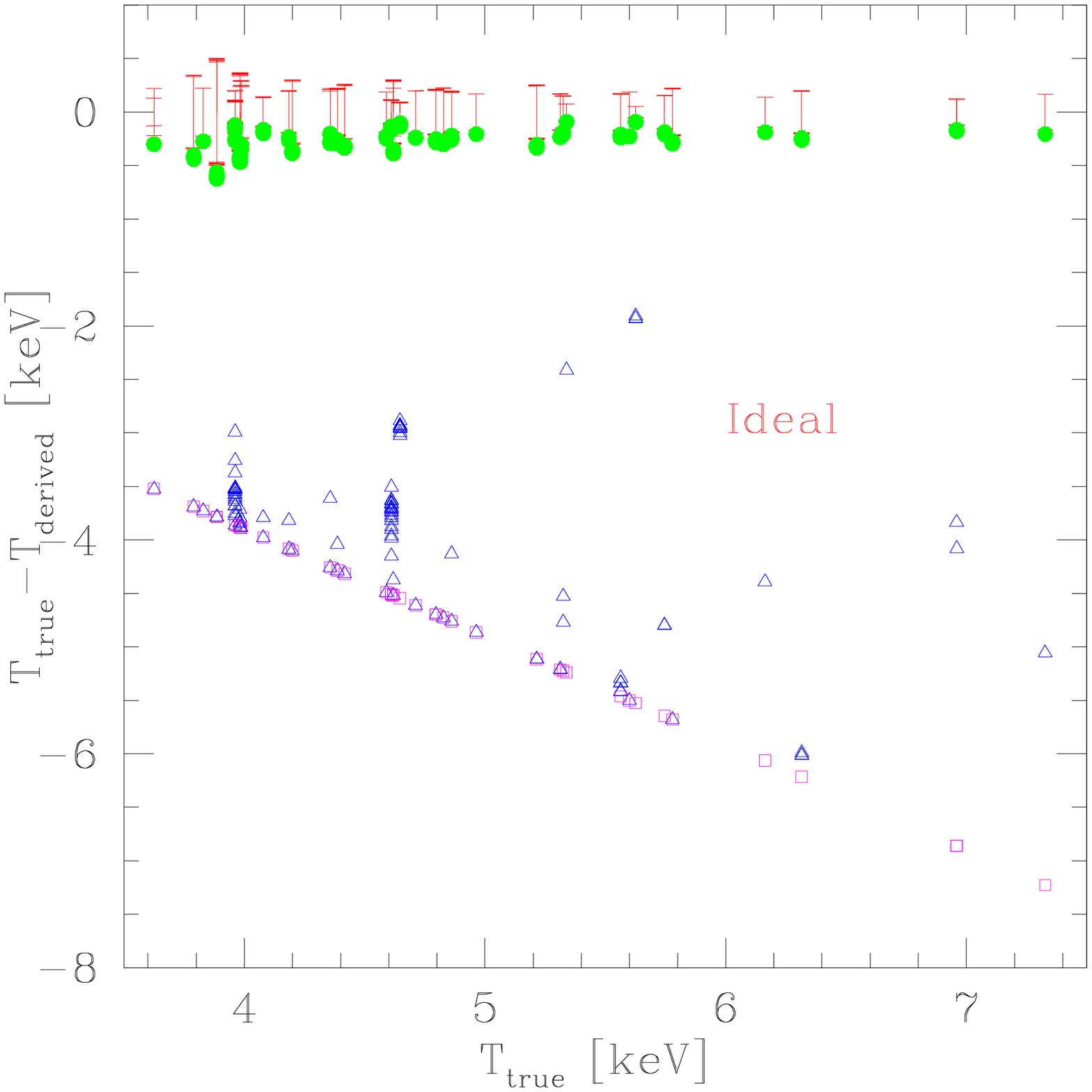,width=0.8 \textwidth} 
\end{center}
\vspace{-0.8cm}
\caption{Derived electron temperature in keV
for an ideal experiment.  The very small statistical error-bars
are shown in red.  Open triangles (blue) are radio contamination, open
squares (purple) are Poisson+correlated IR galaxies, and filled
circles (green) are for dust fluctuations.}
\label{fig:SPT.con.t}
\end{figure}

The effects of the different contaminations on the SZ parameters are
qualitatively easy to understand. They are illustrated with an
ideal noise free experiment (0.1$\mu$K) observing at 100, 150, 217 and
270 GHz with about 1 arcmin resolution.  
First, a contamination from radio sources
gives a positive contribution at low frequencies, where the dominating
$y$ parameter is determined, leading to a lower value for $y$ 
(see triangles in Fig. \ref{fig:SPT.con.y}). Now, with this lower
value for $y$, the high frequency signal seems too high, which can
only be counterbalanced with a very low temperature (see triangles
in Fig. \ref{fig:SPT.con.t}). Finally, the velocity term reads as
$v_p/T_e$, so with a very low temperature, $v_p$ is also forced to
be very small. In reality, this leads approximately to $T_e \approx
v_p \approx 0$.

Both IR galaxies and dust emission give a positive contribution at
high frequencies. The latter can be compensated for in 3 ways, either 1)
$y$ parameter is larger (see squares in Fig. \ref{fig:SPT.con.y}), 2)
temperature is lower (see squares in Fig. \ref{fig:SPT.con.t}), or
3) peculiar velocity $v_p$ is slightly more negative. We see from
Figs.  \ref{fig:fig.con.y}-\ref{fig:fig.con.vp}, that the actual
result is a rather complicated adjustment between these 3
possibilities. The lines of squares in Fig. 3 appear because of
our quantization in cluster masses and have no physical relevance. The
systematic shift in peculiar velocity for a randomly observed cluster
will be somewhere within the range of the symbols. The contaminating
signal was added at each frequency for each cluster, rather than
a population of contaminating signals; and therefore each cluster is
represented by only one symbol. We thus present an upper limit
to the expected effect of contamination, and the effect for a
given cluster  can be smaller.

In the Planck case, one also has to take the
contamination from the CMB primary anisotropies themselves
 into account. This
contribution and its effect were already studied and evaluated to an
error of a few hundred km/sec \cite[]{haehnelt96,aghanim97}. In
particular, this was done in the context of a Wiener filtering
analysis of the component separation taking a detailed
simulation of the millimeter and submillimeter sky  into account. The
contamination by the primary CMB anisotropies slightly shifts the
central values of the parameters without modifying  the
error-bars significantly. We therefore chose not to display it in the figures.

\section{Future ACT-like SZ surveys}
A number of future experiments are planned to perform SZ surveys of a
hundred to a few thousand square degrees. They expect to detect several
thousand SZ clusters. We investigated the effects of the
contaminations for those future ACT-like surveys that have a narrower
frequency coverage but a smaller beam-size than the Planck
surveyor. 

To be specific, we assumed that the future ACT-like
experiment has three observing frequencies 145, 225, and 265 GHz, with
angular resolution of 1.7 arcmin at 145 GHz (with beam-size scaling as
1/frequency), and sensitivity of 2 $\rm \mu$K for each channel. As a
result of the smaller frequency coverage of such experiments, it will
be more complicated for them to disentangle the SZ effect from
other sources which emit at the same frequencies. They will have a
smaller beam-size of $\sim 1.5$ arcmin as compared to the 5 arcmin
Planck beam. This makes the contribution from interstellar dust
smaller since it is dominated by large scale fluctuations.  The CMB
contamination will be quite negligible due to the sharp cut-off at
small angular scales. On the other hand, those contributions due to
fluctuations of CIB or unresolved radio sources will be larger than in
the Planck case. However, more sources will be resolved; we thus set
the fraction of resolved sources to an optimistic value of 50\%.
Moreover for ACT-like experiments, the number of clusters that fill
the beam is much smaller than for Planck. Up to $z=0.4$ all clusters
with masses between $5.\,10^{13}$ and $10^{16}$ solar masses will be
resolved, i.e. exceed the beam size. The use of adapted filtering
techniques like those proposed in \cite{schafer2004} should help in
reducing the contamination in such clusters. At higher redshifts up to
$z\sim 0.7$, clusters with masses below $2.\,10^{14}$ fill the beam
but they mostly represent the clusters close to the detection limit.
It is only at high redshifts ($z > 1$) that more massive clusters
($1.\,10^{15}$ solar masses) fulfill our condition on the beam.

The few observing channels and moderate sensitivity of an ACT-like
experiment makes determination of temperature and peculiar
velocity very difficult.  Statistical error-bars are typically much
larger than the central values themselves. For temperatures the
smallest statistical error-bars are about 5 keV, and more typical
error-bars are 10-20 keV. For peculiar velocity the smallest
error-bars are several hundred km/sec, and more typically it is
several thousands km/sec. Therefore, even if a given contamination
implies a complete failure in determining the central values, it
still remains within the statistical error-bar.

It is slightly different for the Compton parameter, for which expected
error-bars are between $10$ and $50\%$. Both dust and radio
contaminations will have a negligible impact ($\epsilon_y({\rm dust})
\sim \epsilon_y({\rm radio}) \sim 0.1$). Only IR galaxies will have a
noticeable effect with systematic shifts in $y$ of up to $40\%$ in
the worst cases (and $\epsilon_y({\rm CIB}) = 0.4$).  It has been
suggested~\cite[]{knox2003} that  including a $30$ GHz channel
(from complementary observations) will improve all these
statistical error-bars (in particular for $y$), but as we saw above,
this channel will be significantly contaminated by radio sources,
which again makes $y$ too small (and pushes both $T_e$ and $v_p$
towards zero).

\begin{figure}[htb]
\begin{center}
\psfig{figure=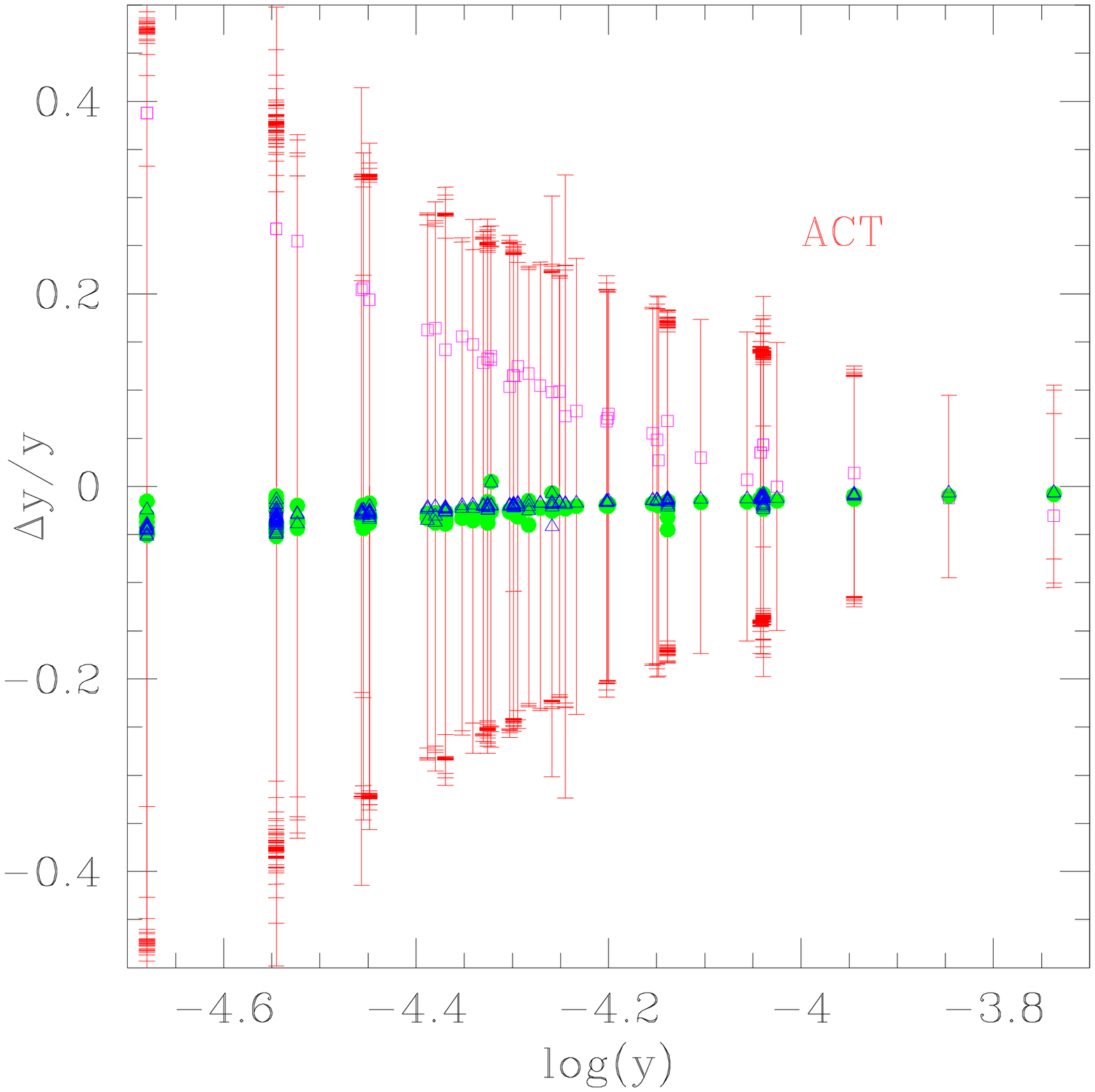,width=0.8 \textwidth} 
\end{center}
\vspace{-0.8cm}
\caption{Compton parameter for ACT-like survey.
The statistical error-bars without contamination are shown in red.
Open triangles (blue) are radio contamination, open squares (purple)
 Poisson+correlated IR galaxies, and filled circles (green) 
dust fluctuations.}
\label{fig:ACT.con.y}
\end{figure}

\begin{figure}[htb]
\begin{center}
\psfig{figure=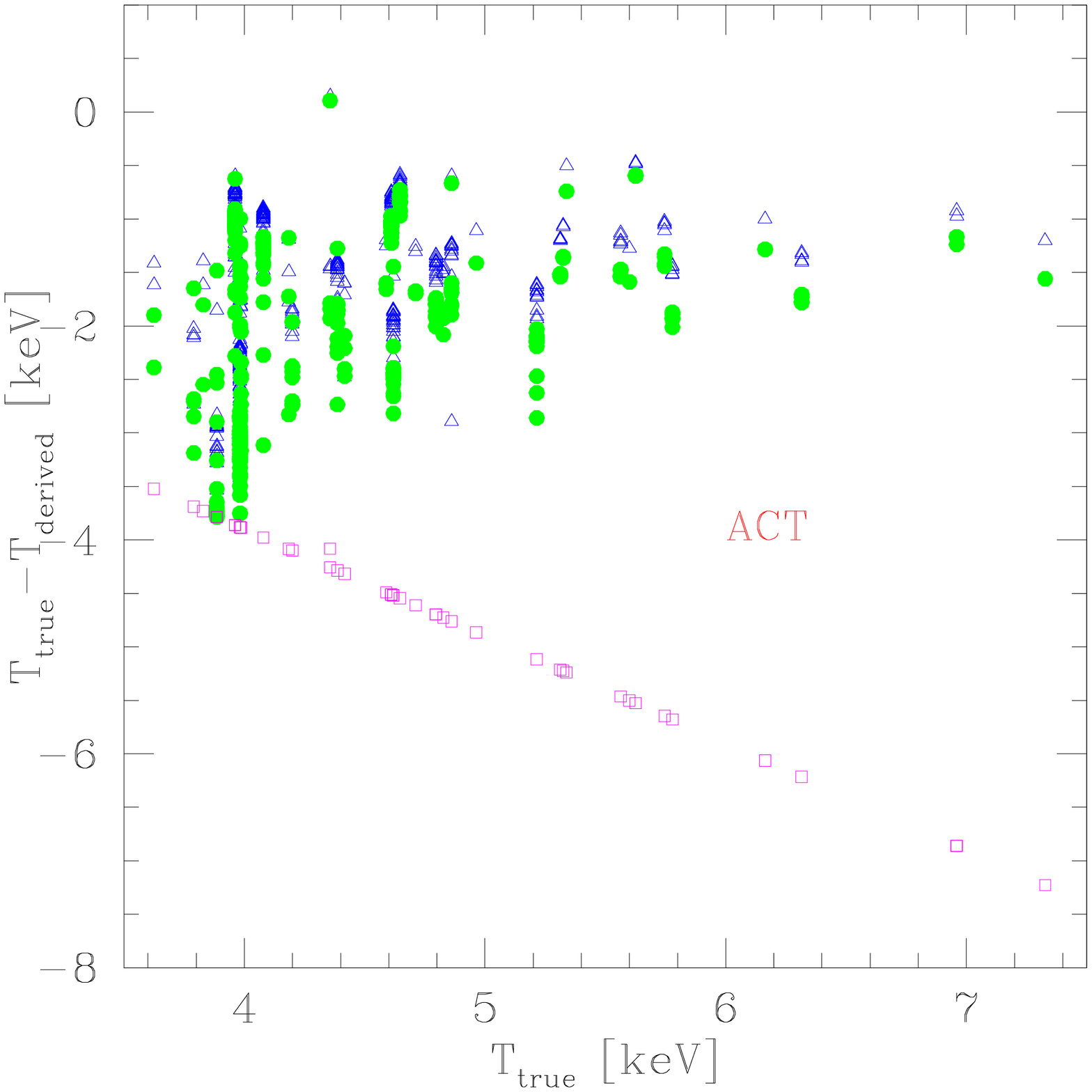,width=0.8 \textwidth} 
\end{center}
\vspace{-0.8cm}
\caption{Derived electron temperature in keV.
Open triangles (blue) are radio contamination, open squares (purple)
are Poisson+correlated IR galaxies, and filled circles (green) are
dust fluctuations. Statistical error-bars (the smallest about 5 keV,
but more typical error-bars are around 10-20 keV) are not shown in
order to avoid confusion.}
\label{fig:ACT.con.t}
\end{figure}

\begin{figure}[htb]
\begin{center}
\psfig{figure=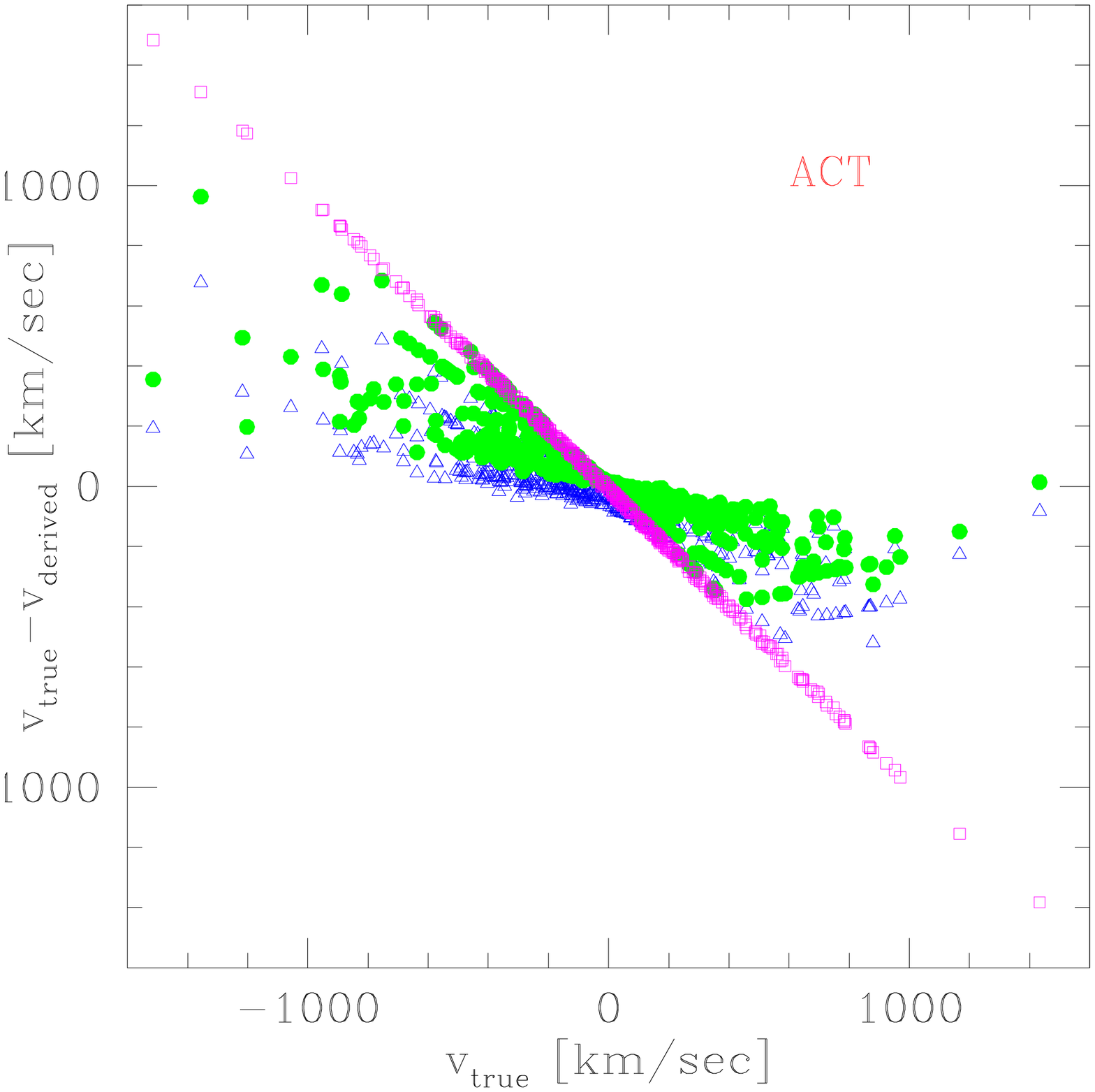,width=0.8 \textwidth} 
\end{center}
\vspace{-0.8cm}
\caption{Upper limit to the expected effect of contamination in
the derived peculiar velocity in km/sec.  Open triangles (blue) are
radio contamination, open squares (purple)  Poisson+correlated IR
galaxies, and filled circles (green)  dust
fluctuations. Statistical error-bars (the smallest are about 500
km/sec, but more typical error-bars are 2000-6000 km/sec) are not
shown, in order to avoid confusion.}
\label{fig:ACT.con.v}
\end{figure}

\section{Requirements for SZ observations}\label{sec:req}

Our results for Planck show that the dust contamination on the 40\%
cleanest part of the sky is not a major source of error for the SZ
parameter estimation even in the most pessimistic case when all
the dust fluctuations remain unremoved. On the other hand, the
fluctuating background from the IR or radio unresolved sources is an
important issue. For the CIB fluctuations, we show that the Poisson
fluctuations do not introduce major systematic shifts in contrast to the
case where we take  the correlations into account. The fluctuations
from unresolved radio sources are the main sources of errors in the SZ
parameters. This contamination dominates over the CIB fluctuations up
to 143 or 217 GHz when only Poisson or Poisson+correlated CIB
fluctuations are present, respectively.

In order to reduce the systematic effects on the SZ parameters, we
need to reduce the contaminations. A rather reasonable demand must be
that for each contaminant the induced systematic error must be smaller
than the statistical error, and hopefully much smaller. If we require
that the ratio of systematic to statistical errors be smaller
than 1/2 for each contaminant, then we find that for Planck the radio
contamination must be reduced by a factor of 13, and the contamination
from CIB must be reduced by a factor of about 5. For the radio sources,
this is attainable when $\sim 60\%$ of the radio background is
resolved in individual sources, as compared to the 12-15\% given by
the present detection limits for Planck. This corresponds to a flux
limit of about 2 mJy at each frequency contaminated by radio emission
(from 3 mJy at 44 GHz to 1.5 mJy at 217 GHz).  For the CIB, 
reduction of the contamination by a factor 5 corresponds to resolving
about 60\% of the background with flux limits ranging between 0.67 and
0.008 mJy from higher to lower frequencies. It is worth noting that
due to confusion limit, at present only 50\% of the CIB is resolved in
individual sources in blank field surveys at 350 GHz
(\cite{lagache2003a}).  This is achieved by the SCUBA instrument in
small fields (less than a few square degrees) and with a 10 arcsec
resolution.

In all cases (for Planck but also for ACT-like), the requirement to
perform SZ parameter estimation from SZ observations is to observe on
regions of the sky where the CIB and radio contamination levels can be
pinned down. This is only possible by resorting to dedicated follow-up
surveys, in radio and IR at higher resolution, to complement the SZ
observations.  However, such follow-up observations, in particular in
radio, have to be done at frequencies close to the SZ measurements to
avoid large errors from the extrapolation of spectral energy
distributions (Ricci et al. 2003). Both large fields (a significant
fraction of the SZ surveys themselves) and angular resolutions (less
than 3 arcsec typically) to beat confusion limits are mandatory.
Future interferometers (e.g. ALMA) will have the required angular
resolution but will not be able to cover large enough areas.
More specifically, for ALMA to resolve 50\% of the CIB in individual
sources (number used for ACT-like experiments) at 230 GHz, the
detection limit is 0.1 mJy ($5\sigma$). In such conditions, 138 days
are needed to survey 1 deg$^2$. It is even worse for resolving 80\% of
the CIB; 96 days are needed for 10 arcmin$^2$ with a detection limit of
0.02 mJy ($5\sigma$). For Planck, only the brightest sources need
to be removed, which can be achieved with rather short observations.

Contamination by radio and IR point sources will be less of a
problem for the interferometric SZ experiments, which naturally have
better resolutions to monitor point sources.

\section{Discussion}
In our study, we have shown that the measurement of SZ parameters ($y$,
$T_e$, and $v_p$), which  is theoretically feasible with multi-frequency
observations of the clusters, is complicated in practice by the presence of
unremoved or non-removable contaminations.  In some cases, it is even
made impossible. Even for very accurate observations of the SZ effect,
i.e. with small error-bars, the contaminations induce systematic
shifts in parameter values. These systematic shifts can be
extremely large, particularly in the case of two sources of
contaminations: fluctuations from unresolved radio sources and
fluctuations of the cosmic infrared background.  Such contaminations
were recently addressed by \cite{knox2003}.  However, these authors
estimated the statistical errors on the SZ parameters rather than the
induced shifts in the derived values. They were also studied by
\cite{white2003}. Rather than studying their effects on the SZ
parameters, these authors estimated the level of noise introduced the
IR and radio sources in SZ surveys, and compared it to the predicted
SZ power spectrum.

For the CIB and radio fluctuations, the systematic shift
dominates the source of errors. Furthermore, the results we obtained can
become even worse, since we chose neither to take  the
amplification of the CIB and radio emission by lensing effects
into account~\cite[]{blain98}, 
nor to include the effects of correlations for the
radio sources  and central radio source in clusters.  Note that
for  ACT-like surveys we took an arbitrarily optimistic level of 50\%
for the unresolved IR and radio backgrounds,  which cannot be
achieved due to the resolution and the frequency of the experiment.

The quantitative results we obtained for the systematic shifts due to
contamination by spurious sources of emission in the SZ frequencies
are based on the hypothesis that the clusters under study fill the
beam of the experiment. This will be the case for the major part of
the Planck cluster catalogue; however, we do not expect this to be the
case for ACT-like survey, for example, as the beam is much smaller ($\sim 1.5$
arcmin). Additional information provided by the spatial extension
of the clusters helps in better constraining the optical depth $\tau$
and thus the Compton parameter itself. Such additional constraints will
reduce the errors on the SZ parameter estimation for the next
generation of experiments like ACT. They will not be as helpful in the
case of Planck, where most of the clusters will fill the
beam. Similarly, measurement of  polarised emission from the
clusters might help in better constraining $\tau$ and thus $y$,
provided the clusters do not fill the beam of the instrument.
However, the polarised signal is weak and the measurement quite
difficult.

It seems from our results that the SZ observations {\it alone} 
will not
provide us with the the three cluster parameters. If we specifically
want to estimate the peculiar velocities, we are forced
to do it on known clusters. It is only on known objects that we might
expect to properly correct for the contaminations as was done
e.g. in \cite{lamarre98} on cluster A2163.  However, such a
procedure will consume time and effort. Another possibility might
be to use extra information from scaling relations and/or X-ray
complementary observations. However, the use of X-ray observed
scaling relations is unwise, partly because the weight from
non-isothermal clusters is different between X-ray observations and SZ
observations (e.g. \cite{hansen04a}), and the use of numerical
scaling relations is still far from being sufficiently reliable due
to the complication in simulating gas dynamics.

\section{Conclusion}
The future blind SZ surveys will be contaminated by astrophysical
sources that can only be partly removed, such as interstellar dust
emission, infra-red galaxies, and radio sources. Such non-removable
contamination will induce a systematic shift in the derived SZ
parameters.  We have shown that IR and radio source-induced systematic
errors may be extremely large, virtually removing the possibility of
measuring peculiar velocity and cluster temperature if using purely the
SZ observations.  Also the systematic shift in the Compton parameter
will be significantly larger than the expected statistical error-bars.
Therefore these contaminations are potentially disastrous for future
survey experiments, and must be considered very as very serious.  On the
other hand, the situation may be less desperate if the contamination
levels are reduced by complementary follow-up surveys.  As mentioned
in Sect. \ref{sec:req}, the radio follow-ups should be planned at, or
close to, the SZ frequencies and the radio source variability requires
simultaneous monitoring. CIB fluctuations need much
better resolution than what is accessible now. One must therefore be
realistic about future SZ surveys, and consider them as excellent
tools for identifying high redshift clusters and not as tools for
measuring peculiar velocity or cluster temperature.

\begin{acknowledgements}
The authors wish to thank an anonymous referee for his/her comments
and suggestions which helped us to improve this study. It is a
pleasure to thank L. Toffolatti for providing us with the radio-source
counts used in the present study. We also thank F.-X. D\'esert and
J.-L.  Puget for useful remarks. This project was partly supported by
the CNRS grant ACI-Jeunes chercheurs ''De la physique des hautes
\'energies \`a la cosmologie observationnelle: d\'eveloppement d'un
groupe de cosmologie sur le campus d'Orsay''.  SHH thanks the Tomalla
foundation for support.
\end{acknowledgements}


\end{document}